\documentclass[aps,draft]{revtex4}
\usepackage[dvips]{graphicx}
\usepackage{amsmath}

\begin{document}

\title{Some Aspects of Statistical Thermodynamics of a Magnetized Fermi Gas}
\author{N.L.Tsintsadze, K.I.Sigua and L.N.Tsintsadze}

\affiliation{ Faculty of Exact and Natural Sciences and Andronikashvili Institute of
Physics, Javakhishvili Tbilisi State University, Tbilisi 0128, Georgia}
\date{\today}

\begin{abstract}

We show that at the Landau ground state a Fermi gas remains precisely a
three-dimensional for an arbitrary magnetic field in radical contrast to the previous claims that the perpendicular component of the pressure of a Fermi gas vanishes at the Landau ground state and therefore, it becomes strictly a one-dimensional gas.

\end{abstract}


\maketitle

Our understanding of the thermodynamics of a Fermi quantum plasma, which is of great interest due to its important
applications in astrophysics [1]-[7] and modern technology [8,9], has recently undergone some appreciable theoretical progress. Answers to some salient questions are given in
Refs. [10-12] with a new type of quantum kinetic equations of the
Fermi particles of various species and a general set of fluid equations.
Having the kinetic equation for the Fermi electron gas a quantum
dispersion equation was derived in the above papers and the
propagation of small longitudinal perturbations were investigated.
Later the dispersion properties of linear oscillations
of quantum electron-ion [13-14] and electron-positron-ion [15,16] plasmas,
as well as of neutral $He^{3}$[17] have been studied.
Whereas the quantum nonlinear
ion acoustic waves (KdV -equation) were investigated in Ref. [14]. The
effects of the quantization of the orbital motion of electrons and the spin
of electrons on the propagation of longitudinal waves, as well as the effect of trapping
in a degenerate plasma in the presence of quantizing magnetic field have
been reported recently [18,19,20]. It should be emphasized that a novel
dispersion relation of a longitudinal wave propagating along a magnetic
field, derived in Ref.[18], exhibits the strong dependence on the magnetic field in radical
contrast to the classical case.

The influence of strong magnetic field on the thermodynamic properties of
medium is an important issue in supernovae and neutron stars, the
convective zone of the sun, the early prestellar period of evolution of the universe. A
wide range of new phenomena arise from the magnetic field in the Fermi gas,
such as the change of shape of the Fermi sphere, thermodynamics, de
Haas - van Alphen [21] and Shubnikov - de Hass [22] effects. Quite
recently an adiabatic magnetization process has been proposed in Ref.[23] for
cooling the Fermi electron gas to ultra-low temperatures.

It should be noted that the diamagnetic effect has a purely quantum
nature and in the classical electron gas it is absent, because in a magnetic
field the Lorentz's force $\frac{e}{c}\overrightarrow{v}\times
\overrightarrow{H}$ acts on a particle in the perpendicular direction to a
velocity $\overrightarrow{v},$ so that it cannot produce work on the
particle. Hence, its energy does not depend on the magnetic field. However,
as was shown by Landau, the situation radically changes in the quantum
mechanical theory of magnetism. The point is that in a constant magnetic
field the electrons, under the action of it, rotate in circular orbits in a
plane perpendicular to the field $\overrightarrow{H}_{0}(0,0,H_{0})$.
Therefore, the motion of the electrons can be resolved into two parts: one
along the field, in which the longitudinal component of energy is not quantized,
$E_{\parallel }=p_{\parallel }^{2}/2m_{e}$ and the second, quantized [24,25], in a plane
perpendicular to $\overrightarrow{H}_{0}$ (the transverse component). Thus, in the non-relativistic case the net
energy of electron in a magnetic field without taking into account its spin
is \ $E(p_{\parallel },l)=p_{\parallel }^{2}/2m_{e}+\hbar \omega _{ce}(l+
\frac{1}{2})$, where $m_{e}$ is the
electron rest mass and $\omega _{ce}=\frac{\mid e\mid H_{0}}{m_{e}c}$ is the
cyclotron frequency of the electron.

If a particle has a spin, the intrinsic magnetic moment of the particle
interacts directly with the magnetic field. The correct expression for the
energy is obtained by adding an extra term $\overrightarrow{\mu }
\overrightarrow{H}_{0}$, corresponding to the energy of the magnetic moment $
\overrightarrow{\mu }$ in the field $\overrightarrow{H}_{0}$. Hence, the
electron energy levels $\varepsilon _{e}^{l,\delta }$ are determined in the
non-relativistic limit by the expression
\begin{equation}
\varepsilon _{e}^{l,\delta }=\frac{p_{\parallel }^{2}}{2m_{e}}+(2l+1+\delta
)\mu _{B},
\end{equation}
where $l$ is the orbital quantum number $(l=0,1,2,3,...),$ $\delta $ is the
operator the $z$ component of which describes the spin orientation $%
\overrightarrow{s}=\frac{1}{2}\overrightarrow{\delta }(\delta =\pm 1)$ and $%
\mu _{B}=\frac{\mid e\mid \hbar }{2m_{e}c}$ is the Bohr magneton.

From the expression (1) one sees that the energy spectrum of electrons
consist of the lowest Landau level $l=0$, $\delta =-1$ and pairs of
degenerate levels with opposite polarization $\delta =1$. Thus each value
with $l\neq 0$ occurs twice, and that with $l=0$ once. Therefore, in the
non-relativistic limit $\varepsilon _{e}^{l,\delta }$ can be rewritten as
\begin{equation}
\varepsilon _{e}^{l,\delta }=\varepsilon _{e}^{l}=\frac{p_{\parallel }^{2}}{%
2m_{e}}+\hbar \omega _{ce}l,
\end{equation}
where $\hbar $ is the Planck constant divided by $2\pi $.

In a series of papers [2-4] Canuto and Chiu considered the thermodynamic properties of a
magnetized Fermi gas and showed that the pressure is different in the
perpendicular and parallel to the magnetic field directions. We note here that this anisotropy
is directly associated with the quantization of energy levels by the
presence of a magnetic field. Canuto and Chiu argued that the perpendicular
component of the pressure $P_{\perp }$ becomes zero at $l=0$. Thus, in their case
there is no lateral pressure and one deals with a physical picture of an
one-dimensional gas.

In this letter, we show that the statement of Canuto and Chiu is false. Namely, at the Landau ground state, i.e. $l=0$, the
perpendicular component of the pressure $P_{\perp }\ $ is not zero. In other
words we will demonstrate that a three-dimensional gases remain three-dimensional
even at $l=0$. To this end, we use the anisotropic distribution function derived
by Kelly [26], also independently by Zilberman [27]. Kelly's result for the
Fermi-Dirac Statistics is the following
\begin{equation}
f_{\alpha }^{k}(\overrightarrow{p}_{\perp },p_{\parallel })=\frac{%
e^{-w_{\alpha }^{2}}}{(2\pi \hbar )^{3}}2\sum_{l=0}^{\infty }\frac{%
(-1)^{l}L_{l}(2w_{\alpha }^{2})}{\exp (\frac{\varepsilon _{\alpha }^{l}-\mu
_{\alpha }}{T_{\alpha }})+1},
\end{equation}
where suffix $\alpha $ stands for the particle species, $w_{\alpha }^{2}=%
\frac{p_{\perp }^{2}}{m_{\alpha }\hbar \omega _{c\alpha }}=\frac{%
p_{x}^{2}+p_{y}^{2}}{m_{\alpha }\hbar \omega _{c\alpha }}$, $\ L_{l}(x)$ is
the Laguerre polynomials of order $l$ [28], for which exist such condition $%
2(-1)^{l}\int e^{-w^{2}}L_{l}(2w^{2})wdw=1$, $\mu _{\alpha }$ is the
chemical potential determined by the normalization condition
\begin{equation}
n_{\alpha }=2\int d\overrightarrow{p}\ \,f_{\alpha }^{k}(\overrightarrow{p}%
_{\perp },p_{\parallel }).
\end{equation}%
Here the factor $2$ is on account of the particle spin.

First, we consider the lowest Landau level, $l=0,\delta =-1$ \ (see Eq.(1)).
In this case the Kelly's distribution function is
\begin{equation}
f_{\alpha }^{k}(\overrightarrow{p}_{\perp },p_{\parallel })=\frac{%
2e^{-w_{\alpha }^{2}}}{(2\pi \hbar )^{3}}\frac{1}{\exp (\frac{p_{\parallel
}^{2}/2m_{\alpha }-\mu _{\alpha }}{T_{\alpha }})+1}.
\end{equation}%
At $T=0$, the Kelly's distribution function (5) reads
\begin{equation}
f_{\alpha }^{k}(\overrightarrow{p}_{\perp },p_{\parallel })=\frac{%
2e^{-w_{\alpha }^{2}}}{(2\pi \hbar )^{3}}{\Large H}\left( \mu _{\alpha
}-p_{\parallel }^{2}/2m_{\alpha }\right) ,
\end{equation}%
where $H(x)$\ is the Heaviside step function and $\mu_\alpha =$\ $\frac{p_{F}^{2}}{%
2m_{\alpha }}.$ Substituting distribution function (6) into Eq.(4) we
obtain the expression of density
\begin{equation}
n_{\alpha }=\frac{m_{\alpha }\hbar \omega _{c\alpha }p_{F}}{\pi ^{2}\hbar
^{3}},
\end{equation}
which is true for the
Lowest Landau level $(l=0)$, i.e. this expression is associated with the
Pauli paramagnetism and self-energy of particles. If we suppose that the
density of electrons is constant, then from Eq.(7) follows an important
statement, namely, that the Fermi momentum decreases along with the increase of a magnetic field. So that a pancake configuration of the Fermi energy thins.

We now derive  the perpendicular component of the pressure using the distribution function (6) for electrons
\begin{equation}
P_{\perp e}=\frac{2}{3}\ 2\int d\overrightarrow{p}\frac{%
(p_{x}^{2}+p_{y}^{2})}{m_{e}}\ f_{e}^{k}(\overrightarrow{p}_{\perp
},p_{\parallel }).
\end{equation}

After a simple integration of  Eq.(8), we obtain
\begin{equation}
P_{\perp e}=\frac{2}{3}\hbar \omega _{ce}\ n_{e},
\end{equation}%
where $n_{e}$ is the density of electrons defined by Eq.(7).

At the temperatures lower than the degeneracy temperature,
$T_{F}=\gamma \left( \frac{n}{H_{0}}\right) ^{2}\ $ (where $\gamma =%
\frac{\pi ^{4}\hbar ^{4}c^{2}}{2m\cdot e^{2}}$) [18], from Eq.(4) and Eq.(5) for the
density of electrons follows such expressions

\begin{equation}
n_{e}=\frac{m_{e}\hbar \omega _{ce}\ P_{F}}{\pi ^{2}\hbar ^{3}}\left\{ 1-%
\frac{\pi ^{2}}{24}\left( \frac{T}{T_{F}}\right) ^{2}\right\} .
\end{equation}%
In this case $n_{e}$ in Eq.(9) is governed by Eq.(10).

Its obvious from  Eq.(9) that at $l=0,\ P_{\perp }$ is not zero and therefore, the Canuto-Chiu
statement is false.

Next, for the parallel component of the pressure, in the same case, i.e. $l=0$ and $T=0$,  we obtain
\begin{equation}
P_{\parallel e}=\frac{1}{3}\cdot 2\int d\overrightarrow{p}\frac{p_{\parallel
}^{2}}{m_{e}}f_{e}^{k}.
\end{equation}
Use of Eq.(6) in Eq.(11) yields
\begin{equation}
P_{\parallel e}=\,\propto \left( \frac{n_e}{{\large H}}\right) ^{2}n_e,
\end{equation}
where \ $\propto \,=\left( \pi ^{4}\hbar ^{4}c^{2}\right) /\left(
9m_{e}e^{2}\right) $.

Finally, we calculate general expressions of the perpendicular and parallel
components of the pressure. For this purpose, we employ the Kelly's distribution
function (3) and derive the expression \ $P_{\perp
e}=\frac{2}{3}\cdot 2\sum_{l=0}^{\infty }\int d\overrightarrow{p}\frac{%
p_{\perp }^{2}}{m_{e}}f_{e}^{k}$ \ or
\begin{equation}
P_{\perp e}=\frac{1}{3}\ \frac{\left( m_{e}\hbar \omega _{ce}\right) ^{2}%
}{\pi ^{2}\hbar ^{3}m_{e}}\sum_{l=0}^{\infty }\ 2 \int_{0}^{\infty
}dw\cdot w^{3}e^{-w^{3}}L_{l}\left( 2w^{2}\right) \int_{-\infty }^{\infty }%
\frac{dP_{\parallel }}{\exp (\frac{\varepsilon _{e}-\mu }{T})+1}.
\end{equation}%
After integration over $w$, \ we get
\begin{equation}
P_{\perp e}=\frac{2}{3}\ \frac{\left( m_{e}\hbar \omega _{ce}\right) ^{2}%
}{\pi ^{2}\hbar ^{3}m_{e}}\sum_{l=0}^{\infty }\ \left( 2l+1\right)
\int_{0}^{\infty }\frac{dP_{\parallel }}{\exp (\frac{\varepsilon _{e}-\mu }{T%
})+1}\ ,
\end{equation}%
where $\varepsilon _{e}$ is defined by Eq.(2). Clearly, at $%
l=0$ and $T=0$ from Eq. (14) follows the result (9).

In the temperature limit $T\ll \mu =\varepsilon _{Fe}$ and $\varepsilon _{Fe}\gg
\hbar \omega _{ce},$ the spectrum is almost continuous, and a quasi
classical approximation is applicable. Thus, in the limit $\varepsilon
_{Fe}\gg \hbar \omega _{ce}$ the maximum quantum number $l_{\max
}=\varepsilon _{Fe}/\hbar \omega _{ce}$ is very large and we can replace the
summation in equations (4), (8) and (11) by an integration $
(\sum_{l=1}^{l_{\max }}\rightarrow \int_{1}^{l_{\max }}dl)$ to obtain
expressions of the density, the perpendicular and parallel components of
the electron pressures.

Introducing dimensionless parameters $\eta _{0}=\hbar \omega _{ce}/\varepsilon
_{F_{0}}$ and $\gamma =\varepsilon _{F}/\varepsilon _{F_{0}}$, from
Eqs. (4) and (14) we obtain
\begin{equation}
n_{e}=\frac{P_{F}^{3}\gamma ^{-3/2}}{3\pi ^{2}\hbar ^{3}}\left\{ \frac{3}{2}%
\eta _{0}\gamma ^{1/2}+\left( \gamma -\eta _{0}\right) ^{3/2}\right\}
\end{equation}
and
\begin{equation}
\frac{P_{\perp e}}{P_{\perp 0}}=\frac{15}{8}\left\{ \eta _{0}^{2}\gamma
^{1/2}+\frac{2}{3}\eta _{0}\left( \gamma -\eta _{0}\right) ^{3/2}+\frac{4}{15
}\left( 2\gamma +3\eta _{0}\right) \left( \gamma -\eta _{0}\right)
^{3/2}\right\} + \\
\frac{15}{16}\,\pi ^{2}\left\{ \left( \frac{T}{\varepsilon _{F_{0}}}\right)
^{2}\left( \frac{1}{\gamma ^{2}}+\frac{2}{\gamma ^{5/2}}\sqrt{\gamma -\eta
_{0}}\right) \right\} ,
\end{equation}
where $\varepsilon _{F_{0}}$ is the Fermi energy at $T=0$ and $\eta_0=0$, $\
P_{\perp 0}=\frac{2}{3}\frac{\left( 3\pi ^{2}\right) ^{2/3}\hbar
^{2}n_{e0}^{5/3}}{5m_{e}}$ is the perpendicular component of the pressure in
the absence of the magnetic field $\eta_0=0$ and at $T=0$.

In order to explicitly express $P_{\perp }/P_{\perp 0}$ through the magnetic field $%
\overrightarrow{H}_{0}(0,0,H_{0}),$ it is necessary to establish the
relation between $\eta _{0}$ and $\gamma .$ To this end, we use the
expression of the Fermi energy
\begin{equation}
\varepsilon _{Fe}=\frac{P_{F}^{2}}{2m_{e}}=\frac{\left( 3\pi ^{2}\right)
^{2/3}\hbar ^{2}n_{e}^{2/3}}{2m_{e}\left\{ \frac{3}{2}\eta +\left( 1-\eta
\right) ^{3/2}\right\} ^{2/3}}.
\end{equation}%
Noting the relation $\eta =\eta _{0}/\gamma $ from Eq. (17), we get
\begin{equation}
\gamma =\frac{\varepsilon _{Fe}}{\varepsilon _{F_{0e}}}=\gamma \frac{\left(
\frac{n_e}{n_{e0}}\right) ^{2/3}}{\left\{ 3/2\eta _{0}\gamma ^{1/2}+\left(
\gamma -\eta _{0}\right) ^{3/2}\right\} ^{2/3}}.
\end{equation}%
If the density is constant $n_{e}=n_{e0},$ Eq.(18) reduces to
\begin{equation}
3/2\eta _{0}\gamma ^{1/2}+\left( \gamma -\eta _{0}\right) ^{3/2}=1.
\end{equation}

Numerical investigation of Eq.(16) at $T=0$ along with Eq.(19)
for various magnetic field shows that there are two distinct regimes of
anisotropy of the pressure via magnetic fields. It can be seen from Figure 1,
where normalized perpendicular component of the pressure as a function of a magnetic
field is depicted, that the behavior of graph on the left and right
sides of the critical point, i.e. $\hbar \omega _{c}=\varepsilon _{F}$ \ $%
(\eta _{00}=0.7631428)$ are quite different. Namely, on the left side of the
critical point the pressure \ $P_{\bot }$ increases along with the increase of
the magnetic field and then decreases towards the critical point to $\frac{%
P_{\perp }}{P_{\perp 0}}=0.9539.$ Rather different process is observed on the
right side of the critical point. Namely, the pressure linearly increases along with the increase of the magnetic field.
We also note here that the critical point is displaced towards larger magnetic fields when density is increased.

We now derive the parallel component of the electron pressure $P_{\parallel e
}=\frac{2}{3}\  \int d\overrightarrow{p}\  \frac{p_{\parallel }^{2}}{%
m_{e}}f_{e}^{k}$ in the same approximation as for the calculation of $%
P_{\perp e}$. The result is
\begin{equation}
\frac{P_{\parallel e}}{P_{\parallel 0}}=\frac{5}{2}\gamma ^{3/2}\eta
_{0}+\left( \gamma -\eta _{0}\right) ^{5/2}+\frac{5}{16}\pi ^{2}\left( \frac{%
T}{\varepsilon _{F_{0}}}\right) ^{2}\left( \frac{\eta _{0}}{\gamma ^{1/2}}%
+2\left( \gamma -\eta _{0}\right) ^{1/2}\right) ,
\end{equation}%
where $P_{\parallel 0}=\frac{1}{3}\frac{\left( 3\pi ^{2}\right) ^{2/3}\hbar
^{2}n_{e0}^{5/3}}{5m_{e}}.$

In Eqs.(16) and (20) the first terms are the contribution from the
lowest Landau level $\left( l=0\right) ,$ i.e. these terms are associated
with the Pauli paramagnetism and self-energy of particles. The other terms are
the result of summation over all higher Landau levels.

We plot the normalized parallel component of the pressure against
the magnetic field $\eta _{0}$ at T=0 on Figure 2, which shows  that the pressure increases along the magnetic field, then decreases towards the critical point and after this point decreases as a square of the magnetic
field. At the critical point $\eta _{0}=\eta _{00},$ the pressure is \ $\frac{P_{\parallel e}}{P_{\parallel 0}}=1.27.$

It should be noted that Eqs.(16) and (20) with expression (19) describe the thermodynamic properties in
the non-relativistic limit with arbitrary magnetic fields.

Neglecting the temperature correction and in the absence of the magnetic
field $\eta _{0}=0$, the sum $P_{\perp e}+P_{\parallel e}$ reduces to the well
known expression of the pressure
\begin{equation}
P_{e}=P_{\perp e}+P_{\parallel e}=\frac{\left( 3\pi ^{2}\right) ^{2/3}\hbar
^{2}n_{e}^{5/3}}{5m_{e}}
\end{equation}

To summarize, we have studied the thermodynamic properties of a magnetized Fermi gas in
the non-relativistic limit deriving general expressions of the perpendicular and parallel
components of the pressure for an arbitrary magnetic fields. Contrary to a false belief of Canuto and Chiu, who have investigated properties of high-density matter in
intense magnetic fields, that the perpendicular
pressure of the Fermi gas becomes zero at the Landau ground state, $l=0$ and $s=-1/2$, and the Fermi gas becomes
exactly a one-dimensional gas, we have demonstrated that at the Landau ground state the perpendicular component of the pressure is not zero and hence, the Fermi gas remains exactly three-dimensional in any magnetic field. In addition, various regimes of anisotropy of the pressure via the magnetic fields are identified in our numerical analysis.
The results of the present paper offer the prospect of vast number investigations revisiting the papers that have used incorrect assumptions of Canuto and Chiu.

\begin{center}
 Acknowledgments
\end{center}

N.L.T. and K.I.S. would like to acknowledge the partial support of
GNSF Grant Project No. FR/101/6-140/13.

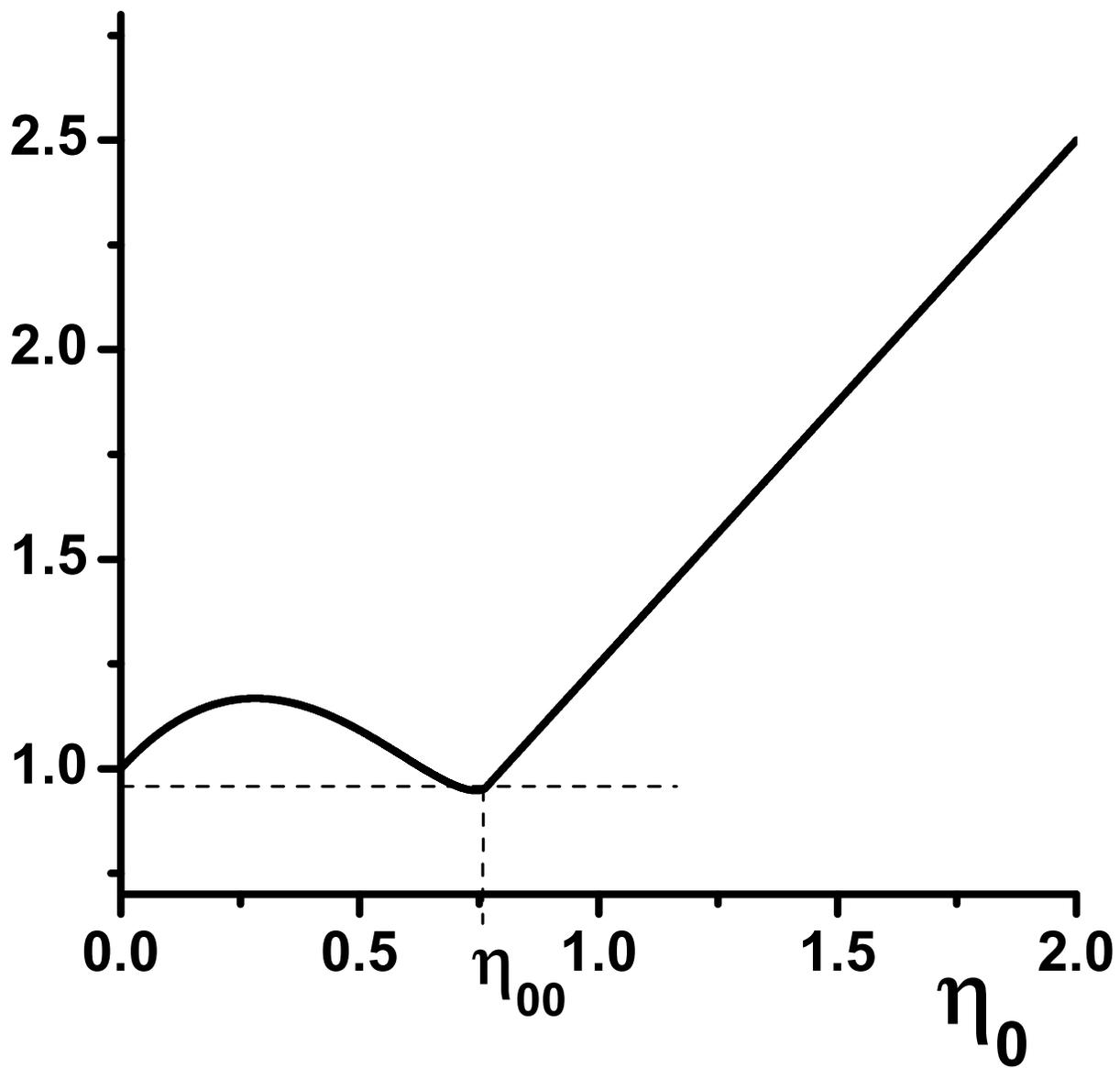
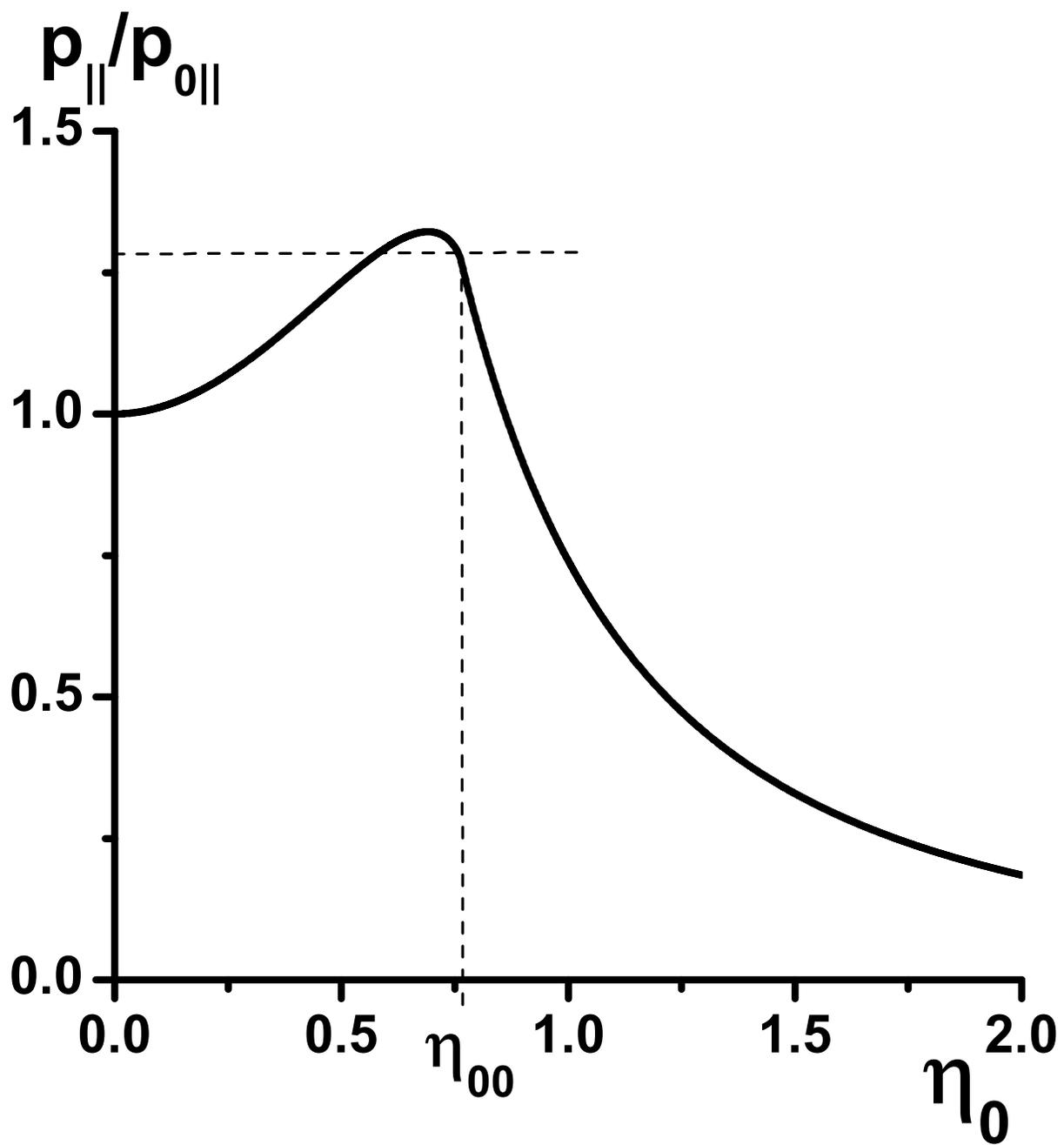
\end{document}